\documentclass[article,preprint,amsmath,amssymb]{revtex4-1}

\usepackage{graphicx}% Include figure files
\usepackage{dcolumn}% Align table columns on decimal point
\usepackage{amsmath}
\usepackage{bm}% bold math

\begin{document}

\title{Realization of free-standing silicene using bilayer graphene}

\author{M. Neek-Amal$^{1}$, A. Sadeghi$^{2}$, G. R. Berdiyorov$^{1}$ and F. M. Peeters$^{1}$}
\affiliation{$^{1}$Departement Fysica, Universiteit Antwerpen,
Groenenborgerlaan 171, B-2020 Antwerpen, Belgium.\\
$^{2}$Department of Physics, Basel University, Klingelbergestrasse 82, CH-4056 Basel, Switzerland.}

\date{\today}

\begin{abstract}

 The available  synthesized silicene-like structures have
been only realized on metallic substrates which are very different
from the standalone  buckled silicene, e.g. the Dirac cone of
silicene is destroyed due to lattice distortion and the interaction
with the substrate. Using  graphene bilayer as a scaffold a novel
route is proposed to synthesize  silicene with electronic
properties decoupled from the substrate. The buckled hexagonal
arrangement of silicene between the graphene layers is found to be
very similar to the theoretically predicted standalone buckled
silicene which is only very weakly van der Waals coupled to the
graphene layers with a graphite-like interlayer distance
of 3.42\,\AA~and without only lattice distortion. We found that these
stacked layers are stable well above room temperature.

\end{abstract}

\maketitle

\textit{Introduction}. ~Silicene, a monolayer of hexagonally
arranged Si atoms,
 has attracted a lot of attention in recent years due to its outstanding physical and
 chemical properties.\cite{Takeda,Roth,Zhang,Guzman}  Due to the similarity of the lattice structures,
 silicene presents essentially the same electronic properties as graphene.\cite{Guzman,Lebegue,Cahangirov} For example,
 the band structure of silicene resembles that of graphene with bands crossing linearly at the Fermi level, resulting in
  a massless Dirac fermion character of the charge carriers. However, unlike graphene, silicene is not stable in a flat
  configuration: silicon's larger ionic radius and therefore larger interatomic distances results in a slightly buckled structure with
   partial sp$^2$ orbitals.\cite{Cahangirov,SahinPRB} On the other hand, such buckling creates new possibilities
   for controlling the band structure of silicene electrically or by functionalizing it with different
   functional
   groups.\cite{Ni,Drummond,Cheng,Spencer,Gao,Quan,peeters}

Since free standing silicene has not been observed in nature and
there is no analogue of graphite in the form of stacked layers of
silicon, increased efforts have been devoted in recent years to
synthesis silicene sheets by implementing more sophisticated
methods. A promising approach is to deposit silicon on metallic
surfaces (e.g. Ag, Ir, Er, ZrB$_2$). However, this resulted into a
plethora of very different crystallographic Si structures depending
on the growth conditions and the particular atomic arrangements of
the substrate
surface.\cite{Lalmi2010,Lin2012,Vogt2012,Feng2012,Chen2012,Jamgotchian2012,Chen2013,Meng2013}
Domain formation, interaction with the substrate and epitaxial
strain have turned out to be determining factors that have prevented
the observation of typical characteristics
 of the theoretically predicted free-standing silicene.

 In this context, Ag surfaces have been successfully used to grow
and synthesize silicene sheets and nanoribbons (see Ref. \cite{Kara}
for review). For example, it was recently demonstrated
theoretically~\cite{WangARXIV} that the Dirac cone of silicene when
epitaxially grown on Ag (111) is destroyed at the $K$-point and
that the experimentally~\cite{Vogt2012,Chen2012} found linear
 dispersion is coming from the Ag substrate. This conclusion agrees with recent Landau level measurements~\cite{Lin2013} of silicene on
 Ag that were supplemented with band structure calculations.
The two
 main  reasons for loosing the Dirac cone in silicene are  lattice distortions and the
 strong chemical  interaction of  silicene with Ag resulting in strongly hybridized states.

Here we present a novel approach where we use graphene layers
to gently confine the silicene layer.  This approach is expected to
result in stable silicene with characteristics that are very close
to those predicted for standalone buckled silicene. We show that the silicene  Dirac cone is preserved and does not interfere with the graphene Dirac point.
 The silicene layer is found to be very different from %those of
silicene on an Ag substrate \cite{Gao_nature}
  and is stable beyond room temperature. Since such structures can naturally arise during epitaxial growth of few-layer graphene on bulk
  silicon carbide (SiC) by thermal decomposition~\cite{Berger2006}, our findings can be useful in the understanding of the mechanisms
  for synthesis of multilayer graphene on SiC.\cite{Singh2011,Xia2012,Wang2012}
   The results may also initiate further research on graphene-silicene stacked heterostructures
   with promising structural and electronic properties.

\textit{Computational method}.~In order to investigate the
stacked heterostructure,  first-principles calculations were
performed in the framework of the density functional theory (DFT)
within the Perdew-Burke-Ernzerhof (PBE) generalized gradient
approximation~\cite{pbe} as implemented in VASP~\cite{vasp}.
Corrections due to the van der Waals interactions are introduced
using the method of Girmme~\cite{Grimme}. A plane-wave basis set
with an energy cutoff of 400~eV was used to expand the valence
electronic wave functions while the projector
augmented-wave method~\cite{paw} was used to treat the cores. A vacuum
layer thinker than 10~\AA\ was used to separate the layers in the
supercell which consists of  ($5\times 5$) silicene unit cells
between two ($8\times 8$) graphene unit cells. The lattice mismatch
for this configuration is less than 2.5~\% based on the separately
calculated lattice constants of silicene (3.838~\AA) and graphene
(2.460~\AA). For such a large supercell, momentum space was
sampled using a $\Gamma$-centered 5$\times$5 mesh for geometry
optimization whereas for calculating  the density of states (DOS) a
7$\times$7 mesh was used. The atomic positions of all atoms were
relaxed until the force on each atom reduced to less than
0.02~eV/\AA.

\begin{figure}[b]
\includegraphics[scale=0.4,angle=-0]{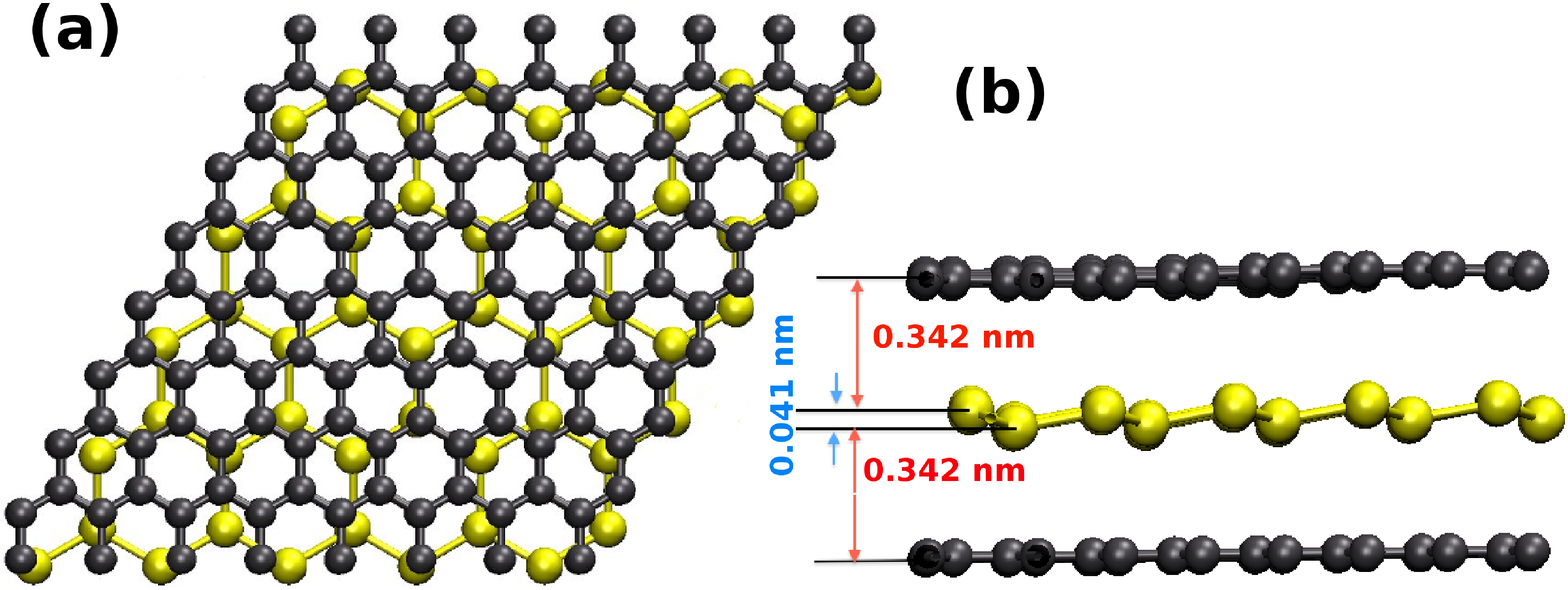}
\caption{\label{dft-relax}  Top (a) and side (b) views (of
$5\times 5$ primitive cells of silicene between two $8\times 8$
primitive cells of graphene layers as relaxed with DFT.
}
\end{figure}

\begin{figure}[b]
\includegraphics[width=.9\textwidth]{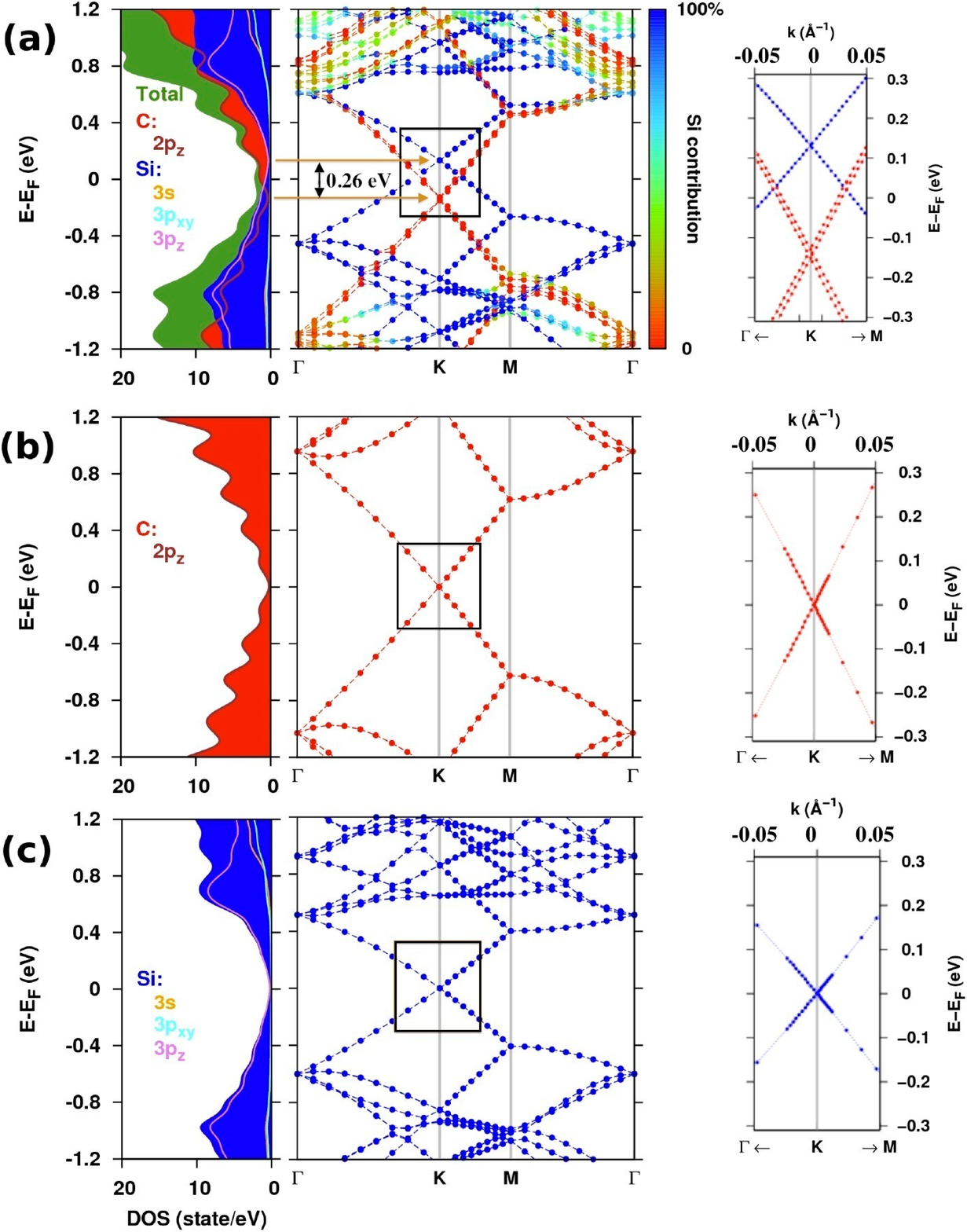}
\caption{\label{dft}  (a) DOS and electronic band
structure calculated for silicene between graphene layers. The band
structure close to the $K$-point is enlarged and shown in the panel
on the right side. (b) and (c): The same as (a) but for pristine
silicene and graphene, respectively. To be comparable, the same
super cell is used in all cases which includes a $5\times 5$ primitive
cell for silicene and a $8\times 8$ primitive cell for graphene.
The relative contribution of silicene is codes by color in the band
structure plots: blue (red) corresponds to the state
originating only from silicene (graphene).  }
\end{figure}

\textit{Stacked silicene-graphene.} The relaxed structure,
depicted in Fig.~\ref{dft-relax} exhibits a 0.41~\AA\
 buckling of the silicene layer and its distance to
each of the graphene layers is 3.42~\AA.
First,  we calculate the DOS for this system using a 49-point grid in  momentum space and
a Gaussian broadening of width 0.1~eV.
 The DOS projected on the spherical harmonics centered
on the atomic positions is then summed up separately over all atoms of the same type
to determine the contributions of separate layers. Only the $2p_z$ orbitals of the carbon atoms
 contribute to the DOS in the shown energy interval.
The contributions of the $3s$, $3p_x$ and $3p_y$ orbitals of the Si
atoms are however vanishing only in the vicinity of the Fermi level.
To compare the projected DOS (PDOS) of the stacked system with those
of the standalone layers, we show in Figs.~\ref{dft}(b) and (c) the
DOS of  standalone graphene and silicene. Instead of using primitive
cells, we used the same supercell as used for the stacked system,
i.e. 5$\times$5 unit cell for silicene and 8$\times$8 unit cell for
graphene; therefore the absolute values of the DOS's of the three cases
can directly be compared
 (the DOS of the freestanding graphene  is multiplied by a factor of two).

Silicene and graphene both are semimetallic with zero
band-gap and vanishing DOS at the Fermi energy. In contrast, the
DOS at the Fermi level is no longer zero  once the layers are
stacked. The minima of the PDOS's of the silicene  and of the graphene
layers do not coincide, and therefore  their sum never becomes zero
%. Between these two minima, the DOS is non-zero (almost constant)
 resulting in metallic behavior of the stacked system. The minima of
PDOS occur at the position of the vertex of the corresponding cone
in the band structure plot as indicated by arrows.
For the stacked system, the relative contribution of the silicene layer
is coded by the color scale, such that a blue (red) point indicates
a predominant silicene (graphene) character. It is clearly seen that the
the Dirac cones of the layers are coaxial but their vertices do not coincide.
The cones corresponding to the two graphene layers are almost degenerate
and the distance between their vertices to the vertex of the
silicene cone along the energy axis is 0.26~eV. This distance is
indeed the same as the difference between the Fermi energies of
standalone graphene (-2.455~eV) and silicene (-2.714~eV).
When the three layers are stacked, their Fermi levels are
shifted with respect to each other so that they become aligned
throughout the system. The fact that the distance between the
vertices of the cones in the interacting system is
the same as the difference of the Fermi energies of the
non-interacting free standing layers implies that the electronic
structure of the stacked layers is almost not affected by
the weak interactions between the layers; the interlayer separation
(3.42~\AA) is indeed too large to cause rehybridization between the
frontier orbitals of the individual layers. Comparing
Figs.~\ref{dft}(a), (b) and (c) shows that, apart from a shift of
the levels, the overall profile of the DOS and the band structure
of the freestanding layers are  preserved in the stacked layers,
in particular close to the Fermi level. This finding is in contrast to
the situation where a silicene layer is put on a metallic substrate, e.g. silicene/Ag
where the silicene layer is strongly deformed resulting in a gap opening
of about 0.3~eV in the silicene layer~\cite{WangARXIV}. The linear
dispersion observed experimentally~\cite{Vogt2012,Chen2012} for this
system has been attributed to the metallic substrate rather than the
silicene layer.

To gain more informative details, we plot the electronic band structures close to the  corner of the Brillouin zone ($K$ point)
for the systems in the RHS of Fig.~\ref{dft}.
In this region where %the Dirac cone
linear dispersion is seen for all three systems.
The slope of the linear fit gives the Fermi velocities as $4.8\times10^5$~m/s  for the silicene branch
and $8.3\times10^5$~m/s  for the graphene branches in the stacked system
%which compares  with the results for freestanding layers as  $3.2\times 10^6$ for silicene and $5.2\times 10^6$~m/s for graphene~\cite{worth}.
in comparison to those of the freestanding layers i.e. $5.2\times 10^5$ for silicene and $8.3\times 10^5$~m/s for graphene.
This confirms the fact that the electronic properties of silicene between double layers of graphene is essentially similar to that of the free standing layers.~\cite{worth}

Notice that several different configurations may be obtained
by laterally sliding the layers with respect to each other. However, performing
DFT calculations for all  possible configurations is  infeasible,
and therefore we focused only on the one that turned out to be the
most stable as determined from molecular dynamics simulations. Since it is
unlikely that in any other configuration the interlayer separation
becomes less than that of the considered arrangement, one expects
that a similar electronic structure is seen for  silicene inside graphene
layers independent of details of the stacking arrangement.

\begin{figure*}[b]
\includegraphics[scale=0.5,draft=false]{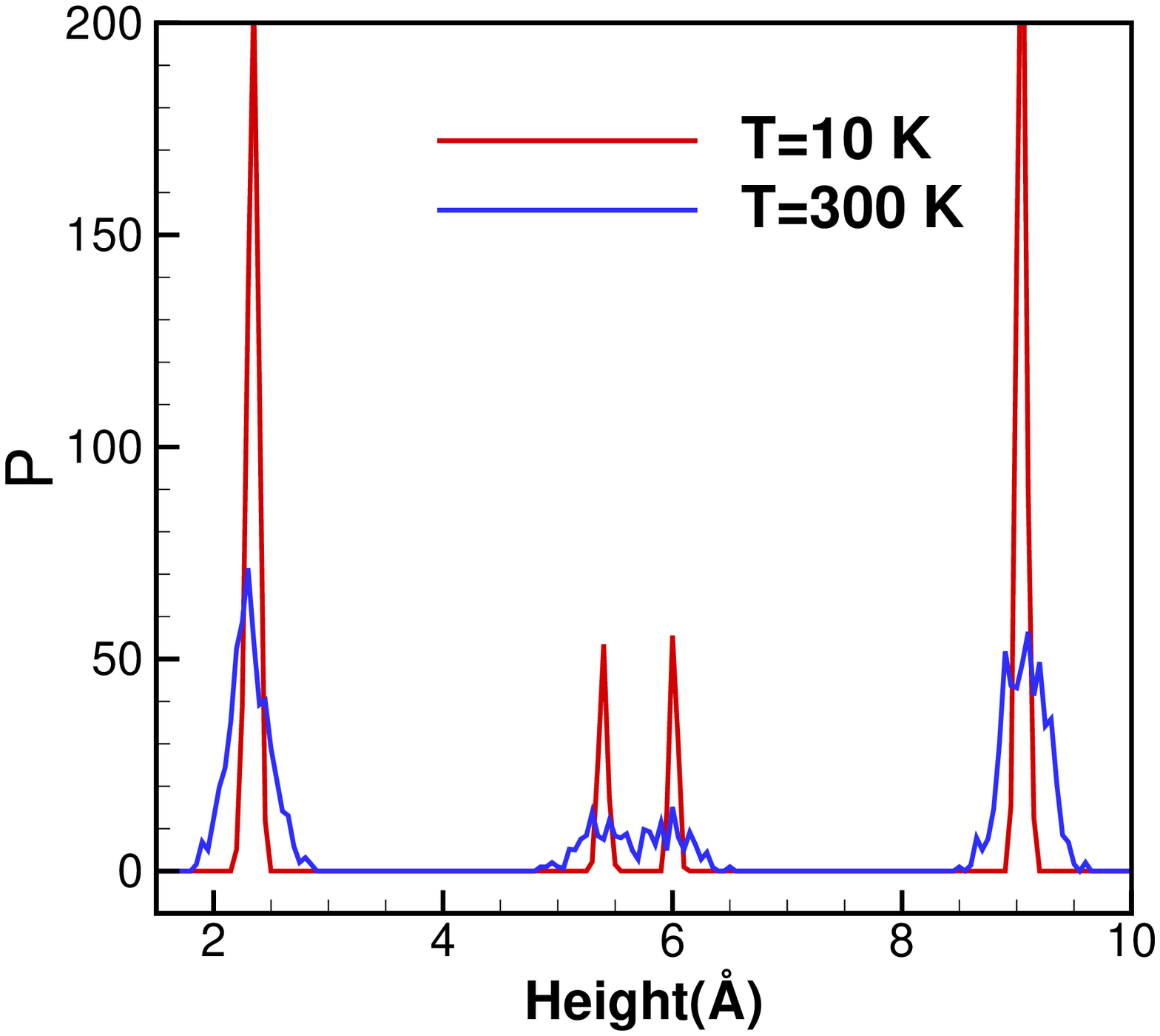}
\caption{\label{MD} \textit{Ab initio} molecular dynamics simulation result for the temperature effect on the structure of stacked graphene-silicene-graphene.
The height distribution of graphene and silicene atoms shows that by increasing temperature the  buckled structure is lost.}
%\caption{\label{MD}  Averaged height distribution of graphene and silicene atoms during \textit{ab initio} molecular dynamics simulation.}
\end{figure*}

{\emph{Thermal stability}. In order to investigate temperature effects on the structure of our proposed confined silicene,
  we performed several \textit{ab initio} molecular dynamics simulations at different temperatures~\cite{DFTB}.
  We used a large sample containing 1224 atoms  for which we were able to observe  ripples.
 The long time molecular dynamics simulations revealed that
  the silicene layer is stable even beyond 1000 K. The graphene layers have ripples due to temperature induced fluctuations.
  The average distance between graphene layers at room temperature is almost identical to that at zero temperature, i.e.  0.72~nm. % $\approx$0.72\,nm.~
   It is interesting to note that the  buckled structure of silicene is affected by temperature
   so that the buckled shape (see Fig. 1(b)) is deformed at room temperature.
For instance, the variance of the buckling height i.e. $\langle h^2 \rangle$  %for the  buckled structure of silicene
at T=10\,K is reduced to 0.033~\AA$^2$ when it is confined between the graphene layers as compared to 0.044\AA$^2$ for the standalone silicene layer.
Figure~\ref{MD} shows the height distribution for T=10\,K and T=300\,K. We found that the in-plane hexagonal lattice structure of silicene preserved at room temperature. Surprisingly we found that the lattice constant of silicene is reduced with about 5$\%$ and it is almost independent of
   temperature up to 1000 K while for each graphene layer both the lattice constant and the  C-C bond lengths are temperature dependent.
    The latter is in contrast to the buckling in fluorinated and hydrogenated graphene~\cite{PRB2013,GA},
     where the buckled structure  remains even at T=1000\,K without  considerable changes in the structures. The reason is
     that in silicene the buckled structure is not due to pure sp$^3$ hybridization while in fluorinated and hydrogenated graphene the
     strong sp$^3$ hybridization plays a determining role in its temperature dependence. The important message is that the here proposed silicene layer at room temperature has a buckling height that
 is random  while the two-dimensional nature is reserved.}

\textit{Conclusions}.~~Using first-principles  calculations and \textit{ab initio} molecular dynamics simulations,
 we investigated the electronic and structural properties and thermal stability of a  silicene layer between bilayer graphene.
Such a silicene layer between graphene layers forms a buckled honeycomb
structure resembling very closely the properties of standalone  silicene.
We demonstrated that  the
electronic and atomic structure of silicene intercalated by graphene layers
is almost identical to the one of  standalone  buckled silicene. Therefore, graphene
layers are an almost ideal template for the formation of silicene. This is in contrast with recently synthesized silicene on top of metallic surfaces where  hybridization modifies
the electronic and structural properties of silicene.

\textit{Acknowledgements} This work was supported by the Flemish Science Foundation (FWO-Vl) and the Methusalem Foundation of the Flemish Government.
  M.N.-A was supported by the EU-Marie Curie IIF postdoc Fellowship/299855.
\newpage

\newpage

\begin{thebibliography}{0}
\bibitem{worth} It is worth noting that when the supercell is composed of $5\times5$
or $8\times8$ (or in general $(3n-1)\times(3n-1)$, where $n$ is an integer) primitive cells, then the corners of the Brillouin zone
are unfolded back again on to the corners of the first Brillouin zone of the primitive cell, i.e. the $K$ points.
Therefore, the Dirac cones of silicene and graphene layers
are coaxial in our case. However, since their vertices are shifted along the
common axis, their intersection (i.e. the region of linear dispersion)
is away from the K-point by $\sim 0.03$~\AA$^{-1}$, as seen in the band structure plot enlarged around the $K$ point in Fig.~\ref{dft}
\end{thebibliography}
\end{document}